\documentclass[prb,floatfix,showpacs,twocolumn]{revtex4}
\usepackage{amsmath}
\usepackage{graphicx}

\begin{document}

\title{From concentration profiles to polymer osmotic equations of state}

\author{C.I. Addison, J.P. Hansen, and A. A. Louis}
 \affiliation{Dept. of Chemistry, University of Cambridge, Lensfield
 Road, CB2 1EW, Cambridge, UK}

\begin{abstract} We show that
 equilibrium monomer and centre-of-mass concentration profiles of
 lattice polymers in a gravitational field, computed by Monte-Carlo
 simulations, provide an accurate and efficient road to the osmotic
 equation-of-state of polymer solutions, via a straightforward
 application of the hydrostatic equilibrium condition.  The method
 yields the full equation of state over a wide range of concentrations
 from a single simulation, and does not suffer from significant finite
 size effects.  It has been applied to self-avoiding walk polymer
 chains with nearest neighbour monomer attractions, from the good
 solvent to the theta solvent regimes.  The consistency of the method
 has been carefully checked by varying the strength of the
 gravitational field.

 \end{abstract}

 \maketitle
\vspace*{-1cm}
 \section[Intro]{Introduction}\label{intro}

 While the osmotic equation of state (e.o.s.)\ of off-lattice models of
polymer solutions or melts can be readily computed from the contact
theorem, according to which the osmotic pressure is proportional to
the monomer density at a hard wall{\cite{Perc76}}, the task is
significantly more arduous for polymers on a lattice.  For short
chains or low polymer concentrations the chemical potential may be
calculated from the insertion probability of a test chain{\cite{Bell73} 
and the pressure then follows by standard thermodynamic
integration.  The insertion method can be extended to solutions of
longer polymers provided configurational bias Monte-Carlo algorithms
are used\cite{Siep92,Frenkelandsmit}.  An alternative method, the 
repulsive wall thermodynamic integration (RWTI) method,
which remains efficient at high polymer concentrations, or in the melt,
was proposed by Dickman\cite{Dick87}. It extends the contact theorem
approach, but requires several simulation runs for increasing
wall-monomer repulsion and subsequent thermodynamic integration for
each state point, rendering the method rather cumbersome. Moreover a
recent analysis by Stukan {\it et.\ al.}\cite{Stuk02}\ revealed that
the RWTI method is prone to large finite-size effects, particularly
at high polymer concentrations, which can only be overcome by
switching to grand-canonical ensemble simulations.

 In this paper we show that the e.o.s.\ of dilute or semi-dilute
polymer solutions is much more efficiently computed by subjecting the
polymers to a strong gravitational field. The resulting sedimentation
profile of the polymer solution then leads directly, via the
hydrostatic equilibrium condition, to the osmotic e.o.s. over a wide
range of concentrations from a single Monte-Carlo (MC) run. The method
extends an idea which has been successfully applied to concentrated
dispersions of rigid colloidal particles (e.g.\ spherical\cite{Bibe93,Piaz93} or plate like\cite{Dijk97}) 
to the case of flexible polymers; it applies to on- and off-lattice models 
alike.

\section{Monomer and Centre-of-mass concentration profiles}

Sedimentation equilibrium of macromolecular solutions or colloidal
dispersions arises from the balance between gravity, which pulls
particles to lower altitudes $z$, and the entropic drive toward
homogeneity, and is characterised by a concentration profile
$\rho(z)$. If $m$ is the buoyant mass of the particles, the
sedimentation length is defined by $\zeta=k_BT/mg$, where $g$ is the
acceleration of gravity and $k_BT$ the thermal energy.  For compact,
micron-sized colloids, $\zeta$ is typically of the order of a few
particle diameters, but for the much lighter fractal polymer chains,
$\zeta$ is very large under normal gravity, so that the solution
remains nearly homogeneous in practice. However, in simulations $g$ can
be tuned to induce a measurable modulation of the concentration
profile.

 We consider systems of N polymer chains, each of L monomers (or
segments) living on a cubic lattice of spacing $a$. Each site hosts at
most one monomer, corresponding to the self-avoiding-walk (SAW) model,
while non-connected nearest neighbour segments interact with energy
$-\epsilon$. The dimensionless inverse temperature is
$\beta^*=\beta\epsilon=\epsilon/k_{B}T ; \beta^*=0$ corresponds to the
athermal SAW limit, while the $\theta$-solvent regime is reached for
$\beta^*\approx0.265$ for L=500 chains\cite{Gras95}.  Each monomer
experiences the gravitational energy $-mgz$, and the gravitational
coupling constant is the dimensionless ratio
$\lambda_{m}=a/\zeta_{m}=mga/k_{B}T$. Sedimentation equilibrium is
characterised by the monomer concentration profile $\rho_m(z)$, {\it
i.e.}\ the mean number of monomers per unit volume at altitude z.  A
more coarse-grained representation focuses on the centre-of-mass (CM)
of each polymer coil, of characteristic dimension equal to the radius
of gyration $R_g$. The gravitational field acts on the CM, and the
corresponding sedimentation length is
$\zeta_{cm}=k_{B}T/Mg=\zeta_{m}/L$ where $M=Lm$ is the total mass of
the polymer. The relevant gravitational coupling constant is
\begin{eqnarray}
\lambda_{cm}=\frac{R_g}{\zeta_{cm}}=\frac{MgR_g}{k_{B}T}=\lambda_m
L\frac{R_g}{a}\sim\lambda_mL^{1+\nu}
\end{eqnarray}
where $R_g\sim a L^\nu$ with $\nu$ the Flory exponent
($\nu\approx 0.59$ in good solvent and $\nu=0.5 $ in $\theta$ solvent).
At sedimentation equilibrium,
the CM concentration profile is $\rho_{cm}(z)$, which satisfies the
normalisation,
\begin{eqnarray}
\int_0^\infty\rho_{cm}(z)dz=n_s,
\end{eqnarray}
where $n_s=N/A$ is the number of polymers per unit area  of an x-y
cross-section of the sedimentation column.

Two limits of the
concentration profile are known explicitly. First,
the system of independent (free) monomers, {\em i.e.}\ $L=1$ polymers,
reduces in the SAW limit ($\beta^*=0$) to a single occupancy lattice
gas in a gravitational field. The corresponding concentration profile
$\rho{_m}{^{(i)}}$ is easily calculated to be:
\begin{eqnarray}
\rho_{m}^{(i)}(z)=\frac{e^{\beta\mu -z/\zeta_{m}}}{1+e^{\beta\mu-z/\zeta_{m}}}
\end{eqnarray}
where $\mu$ is the monomer chemical potential. At sufficiently high
altitudes ($z \rightarrow \infty$) the profile follows the barometric
law for monomers: $\rho_{m}^{(i)}(z) \sim \exp(-\beta mgz) \sim \exp(-
z/\zeta_{m})$. Upon introducing connectivity constraints, however, the
concentration profile contracts significantly. At high altitudes $z$,
where the polymer-polymer interactions are negligible, the
concentration profile takes the form
\begin{eqnarray}\label{barometric}
\rho_m(z) \propto \rho_{cm}(z) \sim exp(-\beta mLgz)\sim
exp(-z/\zeta_{cm}),
\end{eqnarray}
following the barometric law for {\em polymers}.  In other words, it
is contracted by a factor $L$ over a system of unconnected monomers.
This follows from the well-known $1/L$ reduction of the ideal
component of the osmotic pressure of polymer solutions $P^{id} =
\rho_m/L$, where $\rho_m$ is the bulk monomer concentration.  In fact,
if the e.o.s.\ and hence the free energy density of a homogeneous
$(g=0)$ polymer solution is known as a function of bulk polymer
concentration $\rho$ and temperature, the full concentration profile
in the weak modulation limit $\lambda\ll 1$ can be easily calculated
within the local density approximation (LDA)\cite{Bibe93}.
%  However, since only partial information
%on the e.o.s. is available for interacting polymers of finite length
%L, or in the scaling limit L$\rightarrow \infty$~\cite{Gras95},
Here we adopt the opposite point of view. We solve the inverse
problem, and as explained in the next section, we extract the unknown
e.o.s.\ from concentration profiles computed by MC simulations.
\begin{figure}[!htp]
\begin{center}
\includegraphics[width=8cm,angle=0]{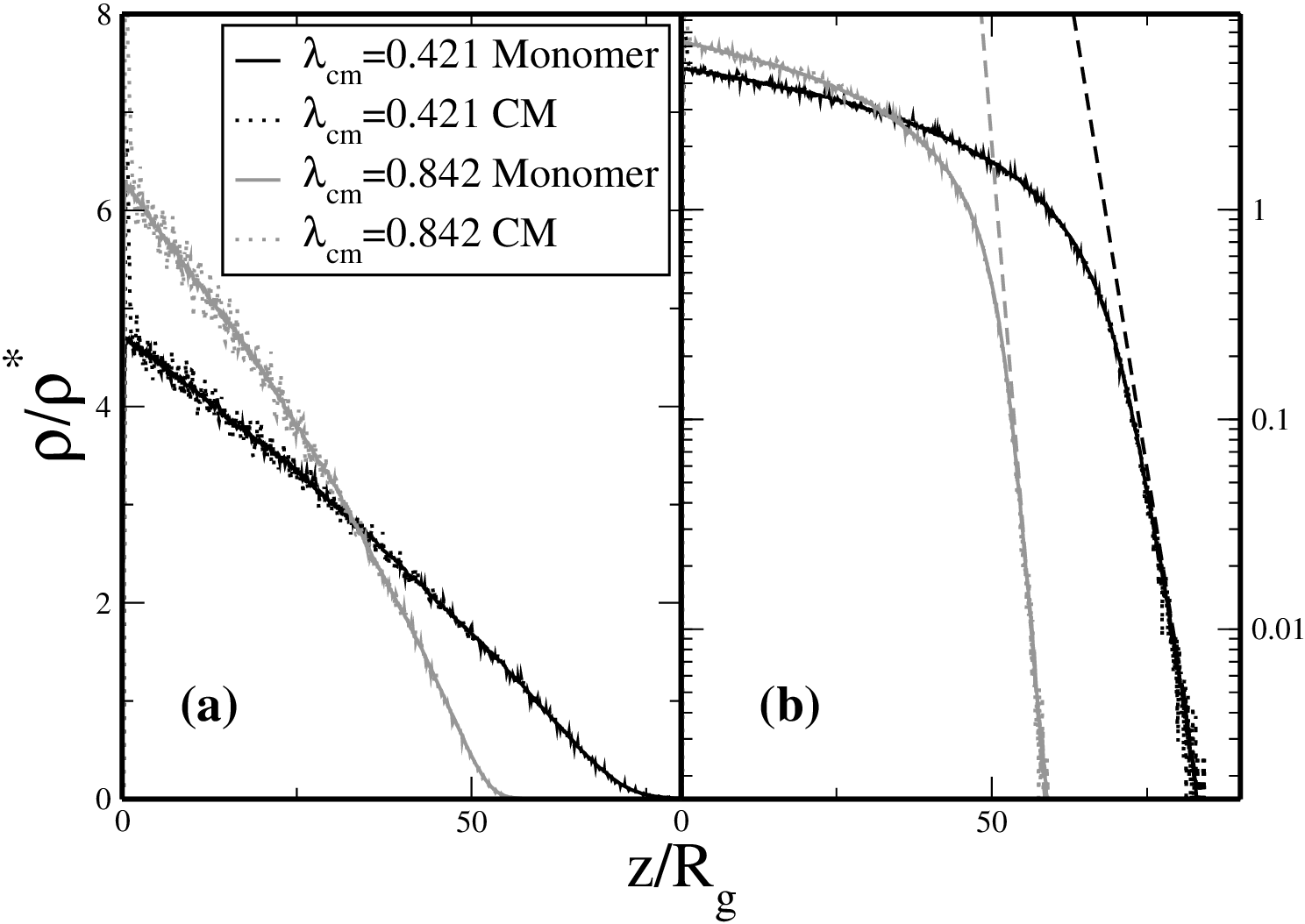}
\caption{ Monomer and centre of mass profiles for $N=1600$ SAW polymer
chains of length $L=500$, shown for two values of the gravitational coupling
constant $\lambda_{cm}$. {\bf (a)} Profiles  plotted on a linear scale. {\bf
(b)} The same
profiles, but now plotted on a logarithmic scale.  The dashed
lines denote the barometric law~(\protect\ref{barometric}), valid at
low densities.}
\label{fig1}
\end{center}\vspace*{-.75cm}
\end{figure}

The simulations were carried out for the cubic lattice model defined
earlier, in a simulation box of dimensions $l_{x}\times l_{y}\times
l_{z}$ in units of $a$.  The square base of area $A=l_{x}\times l_{y}$
was periodically repeated in the x and y directions; most runs were
for $l_x=l_y=100$.  The vertical dimension $l_{z}$ was chosen such
that for reasonable values of the gravitational coupling constant
$\lambda_{cm}$, the polymer concentration at the highest altitude was
negligible compared to the density at the bottom.  Most simulations
were carried out for $N=1600$ polymers of length $L=500$, using pivot,
translation, reptation and configurational bias Monte-Carlo
moves\cite{Frenkelandsmit}. Starting from an initial homogeneous
configuration, the system was equilibrated until the downward drift of
the overall CM of the system stopped.  The profiles $\rho_{m}(z)$ and
$\rho_{cm}(z)$ were then calculated from altitude histograms averaged
over several million configurations (the statistics are of course $L$
times better for the monomer than the CM profiles.)  For given $N$,
$L$ and $\beta^*$, simulations were carried out for several
gravitational couplings $\lambda_{cm}$ Local polymer concentrations
$\rho_{cm}(z)$ are expressed relative to the bulk overlap
concentration $\rho^*=3/4\pi R{_g}{^3}$ where $R_g$ is chosen to be
the temperature dependent radius of gyration at zero concentration.
For a given $N$, higher values of $\lambda_{cm}$ are required to
achieve higher polymer concentrations near the bottom of the
simulation cell.

\begin{figure}[!htp]
\begin{center}
\includegraphics[width=8cm,angle=0]{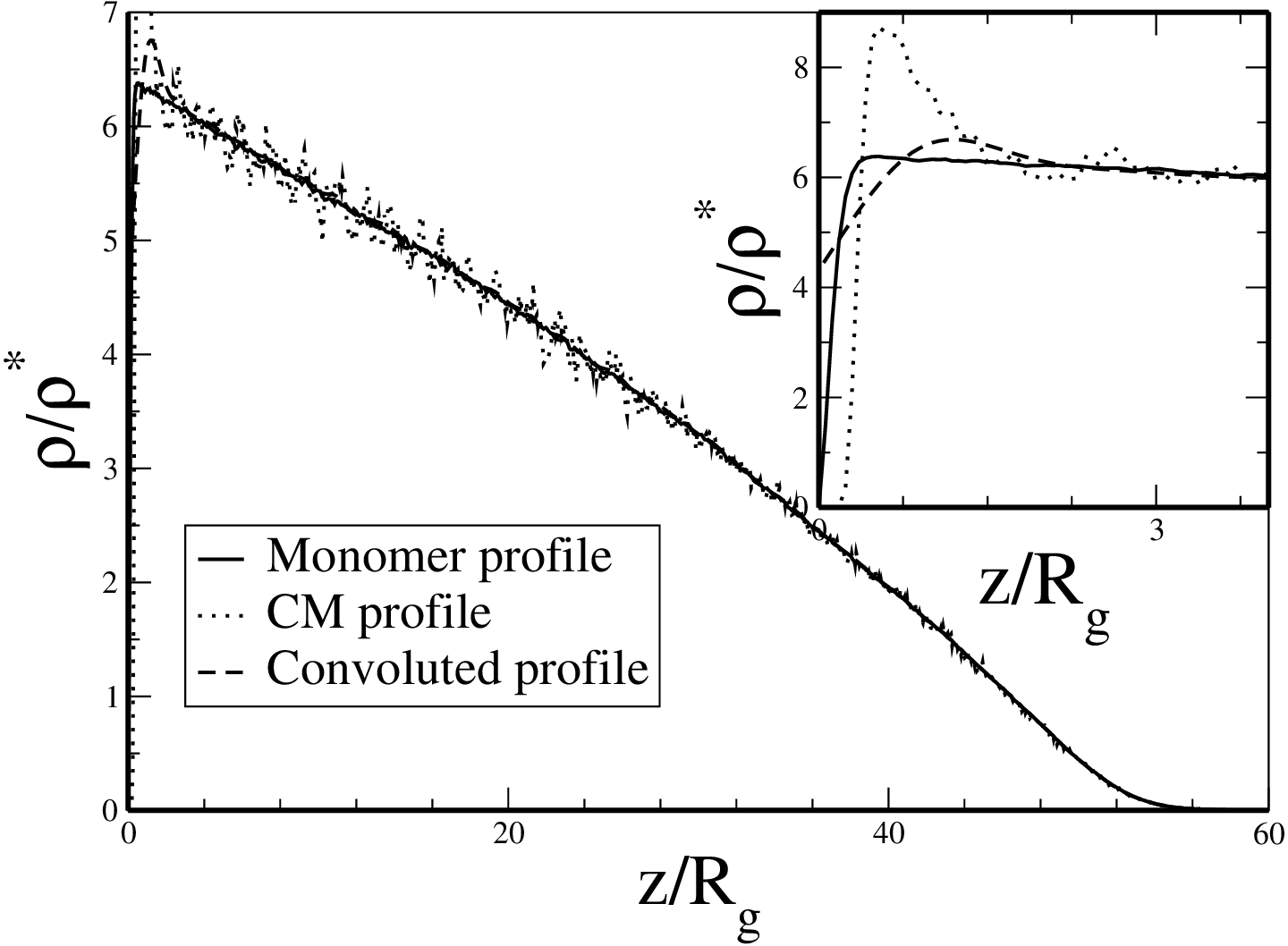}
\caption{Concentration profiles for N=1600 polymers of length L=500, at $\lambda_{cm}$=0.842, and $\beta^*$=0.  
The dashed line is the convoluted profiles obtained from equation (\protect\ref{convol}).}
\label{fig2}
\end{center}\vspace*{-0.75cm}
\end{figure}

Examples of concentration profiles $\rho_m(z)$ and $\rho_{cm}(z)$ for
$\beta*=0$, a model for polymers in a good solvent, are shown in
Figure \ref{fig1}. As expected, the profiles steepen when gravity,
{\it i.e.}\ $\lambda_{cm}$ increases. The CM profiles exhibit a marked
first adsorption layer at the bottom, which is much smaller in
$\rho_m(z)$ profiles~\cite{Bolh01}, but beyond that layer the two
profiles coincide within the statistical errors of $\rho_{cm}(z)$.
The CM layering and preceding ``correlation hole'' in $\rho_{cm}(z)$
(which is more clearly apparent in Figure \ref{fig2}) may be traced
back to the effective wall/CM repulsion of entropic
origin\cite{Bolh01}.  The logarithmic plots of the profiles reveal
linear behaviour at high altitudes with slopes of $-1/\zeta_{cm}$, in
agreement with the asymptotic barometric behaviour, thus providing a
direct check on the convergence of the MC simulations. The close
agreement between $\rho_m(z)$ and $\rho_{cm}(z)$ beyond the CM
adsorption layer is of course a consequence of the polymer
connectivity.  The agreement may be quantified by assuming that the
monomer/CM form factor (which describes the distribution of monomers
around the CM in a polymer coil) is independent of the local polymer
density and taken to be that appropriate for an ideal (Gaussian coil)
polymer, namely:
\begin{eqnarray}
\omega_{CM}(r)=\frac{L}{R{_g}{^3}}e^{(-3r^{2}/2R{_g}{^2})}
\end{eqnarray}
where $r^{2}=x^{2}+y^{2}+z^{2}$ is the monomer/CM distance.  The
approximate relation between $\rho_m(z)$ and $\rho_{cm}(z)$ is then
simply given by the following convolution:
\begin{eqnarray}
\label{convol}
\rho_{m}(z)=\int_0^\infty\omega_{CM}(|z-z'|)\rho_{cm}(z')dz'.
\end{eqnarray}
\begin{figure}[!htp]
\begin{center}
\includegraphics[width=8cm,angle=0]{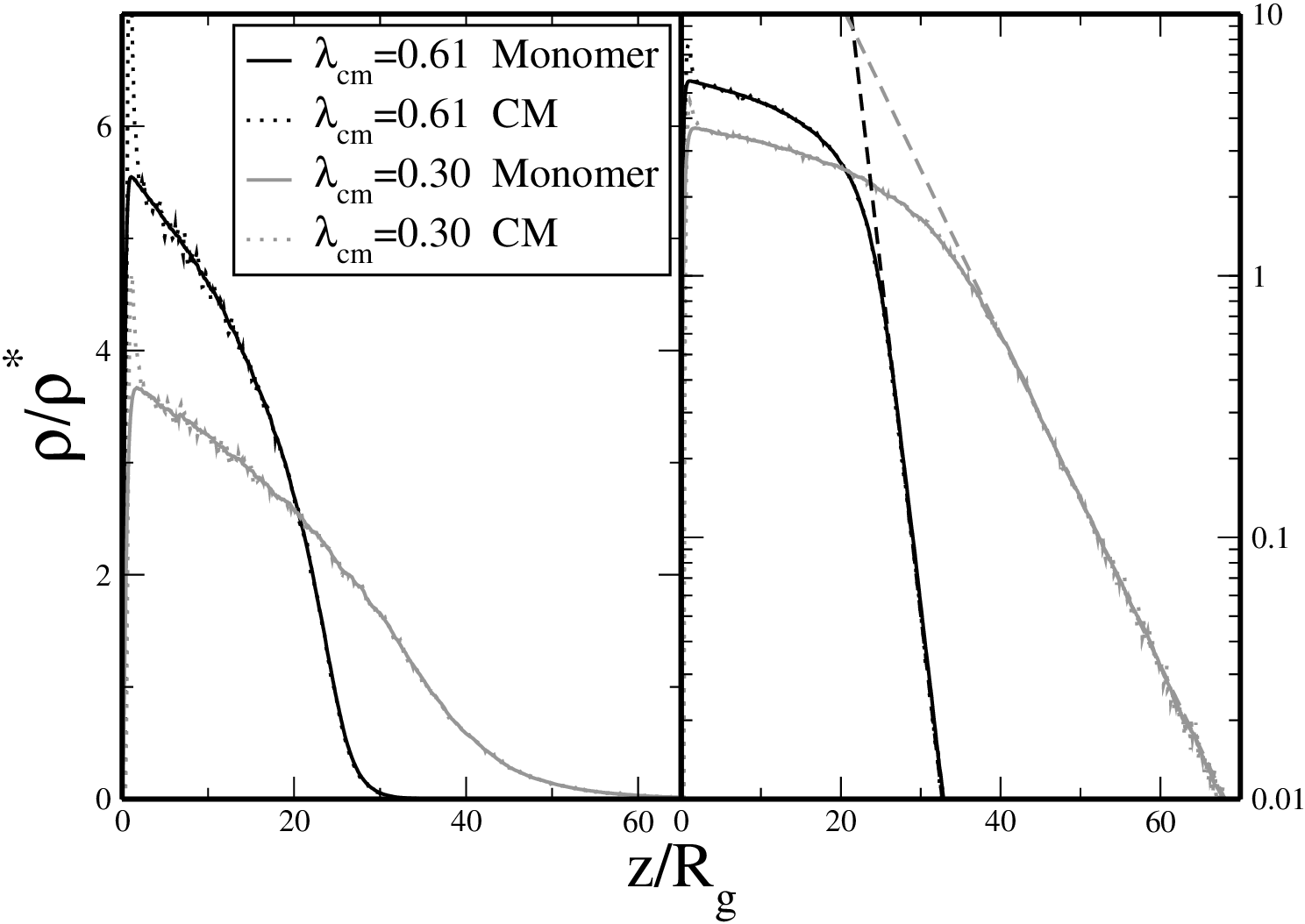}
\caption{Theta solvent monomer and centre of mass concentration profiles. 
In the dilute regime the theta solvent e.o.s.\ resembles that of ideal
polymers, and so follows the barometric law for higher densities than
polymers in good solvent do.}
\label{fig3}
\end{center}\vspace*{-0.75cm}
\end{figure}

An example is shown in Figure \ref{fig2}.  The agreement between the
``exact'' $\rho_m(z)$ and the approximation is seen to be good, except
at distances less than $R_g$ from the wall where the internal
structure of the polymers is expected to be distorted compared to its
bulk behaviour.

Concentration profiles $\rho_m(z)$ and $\rho_{cm}(z)$ under $\theta$
solvent conditions($\beta^*$=0.265) are shown in Figure \ref{fig3} for
two gravitational couplings $\lambda_{cm}$.  The logarithmic plots
show that the profiles reach the asymptotic barometric law earlier 
than in the SAW case, but begin to show important deviations from
the asymptotic limiting law of Eq.~(\ref{barometric}) at $\rho \gtrsim
\rho*$.  In other words polymers behave ideally in a $\theta$-solvent
in the dilute regime $\rho<\rho^*$ only\cite{Addi04}.
\vspace*{-0.5cm}
%The logarithmic plots
%show that the profiles reach their asymptotic ideal solution limit
%significantly faster than under good solvent conditions, but that the
%high density regime($\rho/\rho^*\gtrsim 4$) the profiles deviate
%strongly from ideal solution behaviour and point to rapidly decreasing
%osmotic compressibility of the solution. kIn other words polymers
%behave ideally in a $\theta$-solvent in the dilute regime
%$\rho<\rho^*$ only~\cite{Addi04}.

\section{From concentration profile to osmotic pressure}

In order to extract the osmotic e.o.s.\ from the concentration profiles
of section 2, we proceed as in ref\cite{Bibe93}.  For sufficiently slowly
varying profiles, {\it i.e.}\ for sufficiently weak external field, the
LDA holds and the Euler-Lagrange equation associated with the
minimisation of the free energy functional with respect to the
concentration profile $\rho_{cm}(z)$ leads back to the macroscopic
equation of hydrostatic equilibrium\cite{Bibe93}:
\begin{eqnarray}
\label{eq6}
\frac{dP(z)}{dz}=-Mg\rho_{cm}(z)
\end{eqnarray}
where P(z) is the local osmotic pressure at altitude z. Integration of
Eq. (\ref{eq6}) yields:
\begin{eqnarray}
\label{eq7}
\beta P(z)=\frac{1}{\zeta_{cm}}\int_z^\infty\rho_{cm}(z')dz'
\end{eqnarray}
Thus P(z) and $\rho_{cm}(z)$ are known as functions of altitude, with
elimination of z leading to the desired e.o.s.\ of the bulk polymer
solution $P=P(\rho_{cm}, T)$.
  Note that the LDA approximation is not
expected to be accurate near the hard wall where $\rho_{cm}(z)$ varies
rapidly within the first adsorption layer, so that the lower
integration limit in Eq.~(\ref{eq7}) should be taken at $z\gtrsim R_g$, {\it
i.e.}\ beyond the peak of $\rho_{cm}(z)$, where $\rho_{cm}(z)$ and
$\rho_m(z)$ become indistinguishable.

Eq.~(\ref{eq7}) has an obvious intuitive interpretation: the system
relaxes to a density $\rho(z)$ such that the local pressure
$P(\rho(z))$ counteracts the weight of the polymers above.
Conversely, when the e.o.s.\ is known, Eq.~(\ref{eq6}) can be directly
integrated to calculate the concentration profile\cite{Bibe93}. For 
example, in the semi-dilute regime, where the osmotic equation of state 
obeys the well known scaling law $\beta P(\rho)/\rho \sim \gamma
(\rho/\rho^*)^{1/(3 \nu -1)}$\cite{Corey}, with $\gamma$ a dimensionless 
constant that depends on polymer details, the resulting concentration 
profile is given by:
\begin{equation}\label{eq:semi-dilute}
\rho^{sd}(z) =  \rho_0 \left( 1 -
\left(\frac{\rho_0}{\rho^*}\right)^{\frac{1}{1-3 \nu}}
 \frac{\lambda_{cm}}{3 \gamma \nu }
\frac{z}{R_g}\right)^{3 \nu -1}
\end{equation}
where $\rho_0$ is the maximum polymer density, reached at $z=0$, 
and follows from inverting the equation
\begin{equation}\label{eq:rho_0}
 P(\rho_0) = M g n_s,
\end{equation}
if the  wall-induced layering at the bottom of
the container is ignored.
Eq.~(\ref{eq:semi-dilute}) provides an accurate fit to the density
profile in the semi-dilute regime, but at higher altitude the analytic profile 
$\rho(z)$ fails to cross over to the barometric law as illustrated in Figure~(\ref{fit}).
\begin{figure}[!htp]
\begin{center}
\includegraphics[width=8cm,angle=0]{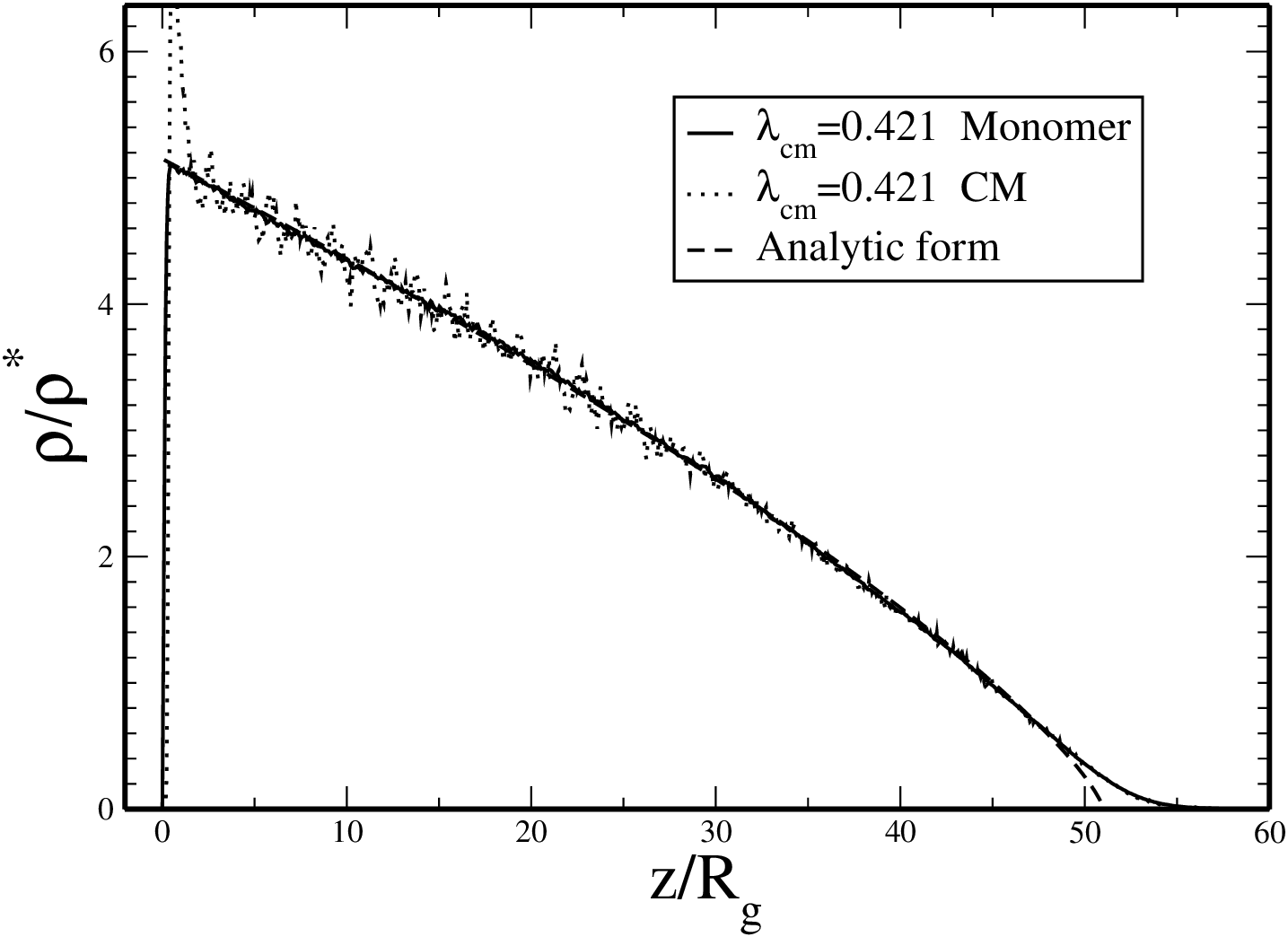}
\caption{Concentration profiles, compared to the analytic form from 
equation (\protect\ref{eq:semi-dilute}).  Note the excellent agreement in 
the semi-dilute regime and the expected disagreement in the dilute regime 
where the barometric law takes over.}
\label{fit}
\end{center}\vspace*{-.75cm}
\end{figure}

 We have carried out the above inversion procedure for $\beta^*$=0, 0.1, 0.2,
and 0.265 ($\theta$ solvent). The resulting e.o.s.\ $P(\rho, T)$ should be
independent of the gravitational coupling used in the MC simulations
provided $\lambda_{cm}$ is not too large, {\it i.e.}\ the
gravitational field is not too strong.  This was checked explicitly by
carrying out the inversion procedure for two different values of
$\lambda_{cm}$; the resulting e.o.s.\ always turned out to be
practically identical, with very small differences (typically less 1\%
over the whole concentration range) providing an estimate for the sum
of systematic and statistical errors.  However, if the applied
gravitational field is too strong ($\lambda_{cm}\gtrsim 1$) the
resulting profiles vary too rapidly with z for LDA to remain accurate.
This reflects itself in the failure of the resulting compressibility
factor Z=$\beta P/\rho$ to go over to its ideal gas limit as
$\rho\rightarrow0$, as illustrated in the inset of Figure \ref{fig4}.
Our results for Z($\rho, T$) are plotted in Figure \ref{fig4}, along
four isotherms, and compared to the predictions of MC simulations of
bulk polymer solutions\cite{Addi04} based on the RWTI
procedure\cite{Dick87}.  The agreement is seen to be excellent
throughout.  In fact some apparent discrepancies between our earlier
data\cite{Addi04}, based on the RWTI method and those obtained with
the present method were resolved when contentious RWTI high concentration 
points were re-run with considerably improved statistics.

 In an effort to detect possible finite size effects, we repeated some
of the simulations with a four times larger base area. Any observed 
differences in the resulting data were within statistical noise.

A related consistency test of the LDA-based inversion procedure is
provided by the scale invariance property of the LDA concentration
profile, according to which, along any isotherm, the re-scaled
profiles $\rho_{cm}(z/\zeta_{cm})$ or $\rho_m(z/\zeta_{cm})$ depend
only on the dimensionless product $n_s\lambda_{cm}a^2$.  This means
that as long as LDA applies, the re-scaled profiles should be
identical if the number of polymers is divided by a given factor,
provided the gravitational coupling $\lambda_c$ is multiplied by the
same factor.  An illustration of this scale invariance is shown in
Figure~\ref{fig5}. The equation-of-state extracted from the density
profiles are in perfect agreement.  Deviations from scale invariance
would signal the breakdown of the LDA assumption, providing a
diagnostic for the consistency of the procedure.
\begin{figure}[!htp]
\begin{center}
\includegraphics[width=8cm,angle=0]{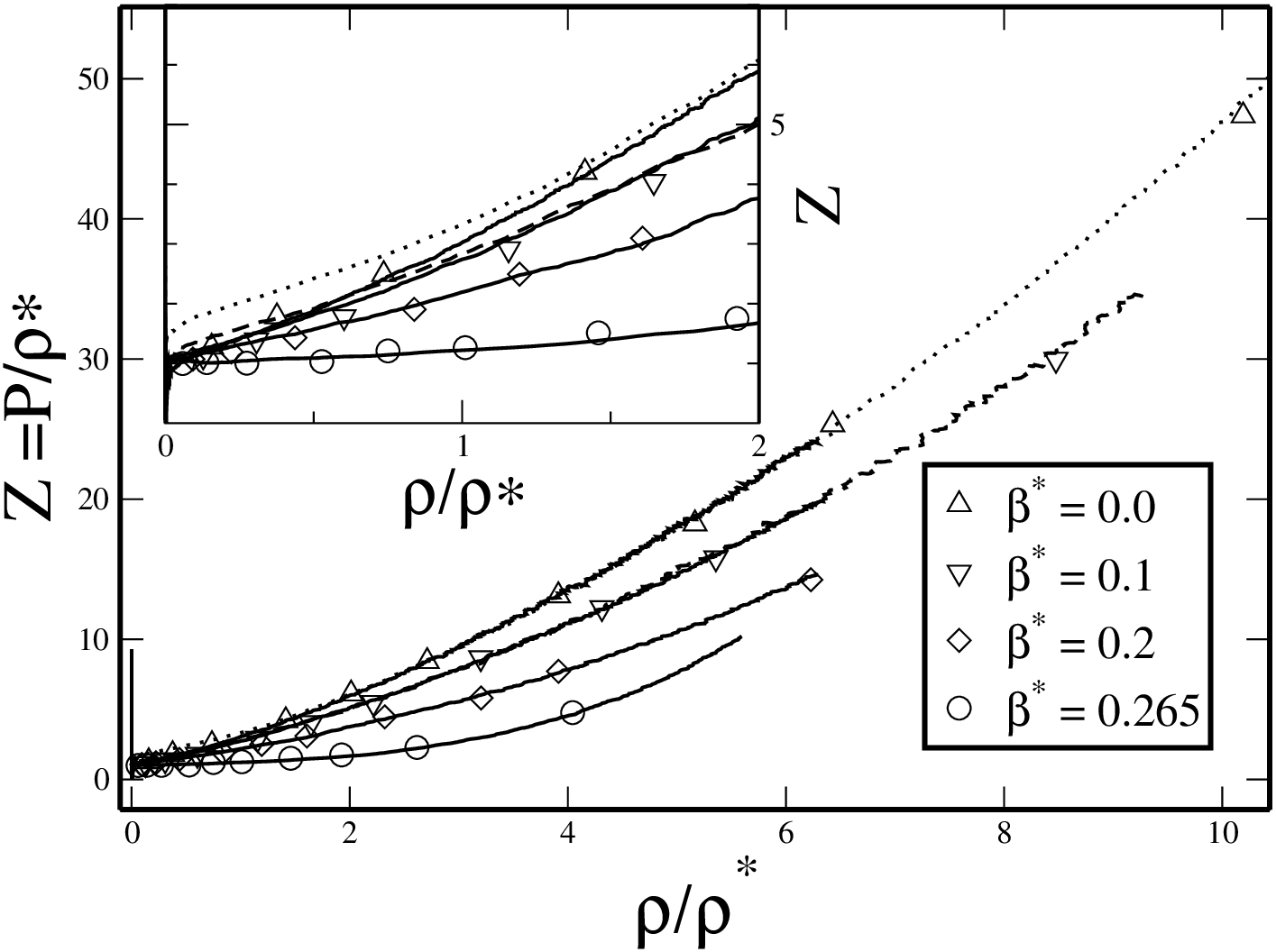}
\caption{Equation of state isotherms calculated by inversion of
concentration profiles (lines) compared to earlier simulations
{\protect\cite{Addi04} using the RWTI method (symbols).  Whereas the
former need about $50$ independent simulations per isotherm, the
present method requires only one simulation per isotherm.  The solid
lines are for runs $\lambda_{cm}<1$, the dotted $\beta^*=0$ line is at
$\lambda_{cm}=4.21$ and the dashed $\beta^*=0.1$ line is at
$\lambda_{cm}=1.97$.}}
\label{fig4}
\end{center}\vspace*{-1cm}
\end{figure}

\begin{figure}[!htp]
\begin{center}
\includegraphics[width=8cm,angle=0]{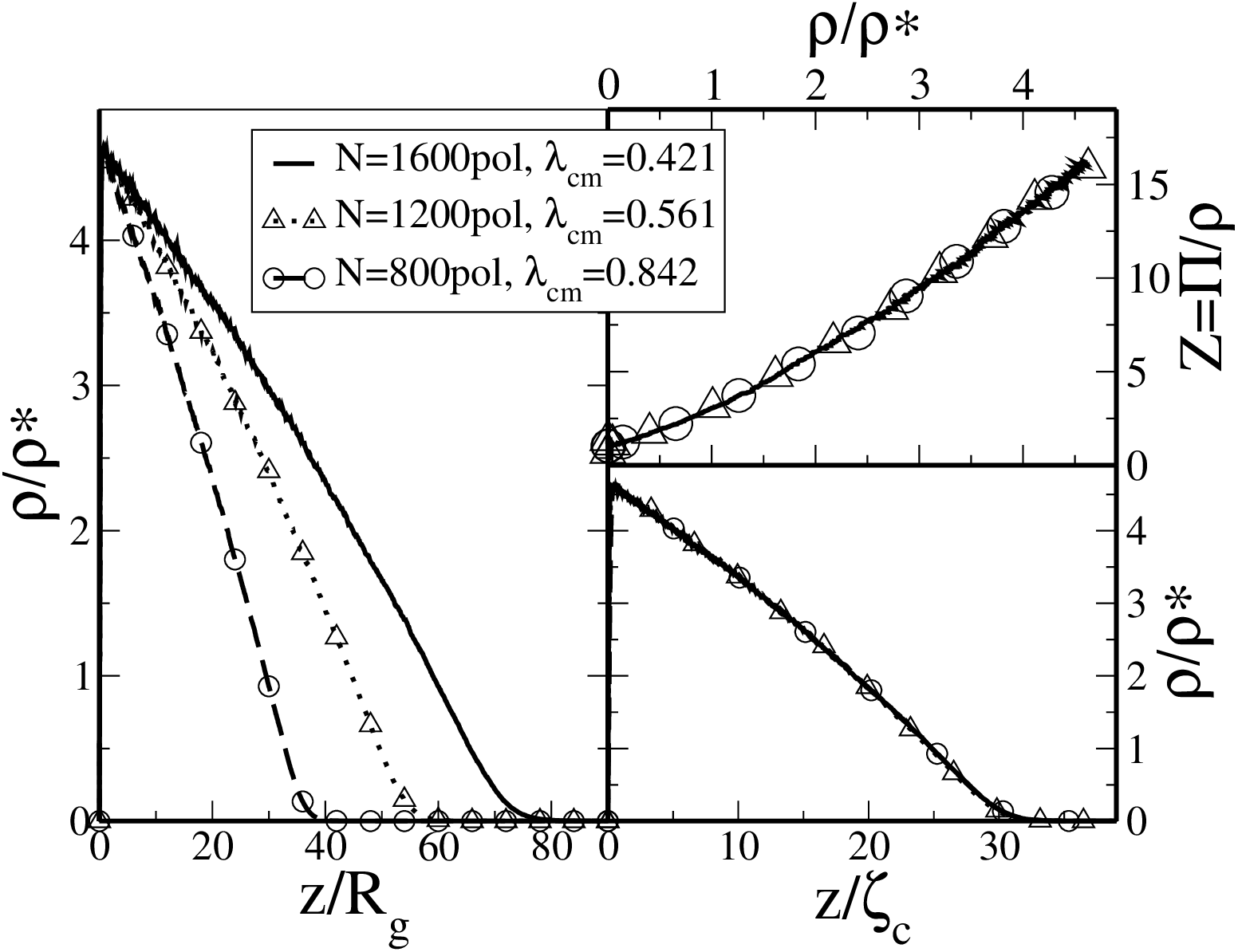}
\caption{{\bf (a)} Three different concentration profiles from simulations 
with different numbers of polymers and gravitational strengths, but with the 
product $n_s\lambda_{cm}a^2$ held constant. Note that the equations of state 
{\bf(b)} generated from the concentration profiles are practically identical, 
as are the concentration profiles {\bf(c)} when scaled by the gravitational length.}
\label{fig5}
\end{center}\vspace*{-.75cm}
\end{figure}

\section{Efficiency of inversion method}

By inverting the concentration profile in an imposed external field, we
measure the full equation of state in a single simulation.  This
contrasts with the RWTI method\cite{Dick87}, where for each value
of $\rho$ a number of separate simulations are needed to perform thermodynamic
integration.  For example, when we calculated the e.o.s.\ of $L=500$
polymers as a function of solvent quality in\cite{Addi04}, using the
RWTI method, we needed about $5$ simulations for each of the $10$
statepoints along each isotherm.  In the current work, we used only
$1$ simulation of similar size to each of the $50$ simulations per
isotherm carried out in\cite{Addi04}. As demonstrated in
Fig.~\ref{fig4}, the results are of comparable accuracy, but achieved
with considerably less CPU time.

For a given $\lambda_{cm}$ and number of polymers $N$, the maximum
density achieved under an external field can be estimated from
Eq.~(\ref{eq:rho_0}).  If the e.o.s.\ scales as $\beta
P(\rho)R_g^3 \sim (\rho/\rho*)^{\alpha}$, then the maximum density  $\rho_0$
scales as
\begin{equation}
(\rho_0/\rho*) \propto \left(\frac{N
\lambda_{cm}}{A/(R_g)^2}\right)^{\frac{1}{\alpha}}
\end{equation}

Larger $\rho_0$ can be achieved by increasing $\lambda_{cm}$, although
this is constrained by the LDA criterion, or by increasing $N$.  For
semi-dilute polymers $\alpha\approx 2.3$ in good solvent and $\alpha
\approx 3$ in a theta solvent\cite{Addi04},\cite{Corey}. To double $\rho_0$, 
the number of polymers $N$ must be increased by a factor five for the
former and eight for the latter solvent quality.  On the other hand,
for the RWTI method, doubling $\rho_0$ simply means doubling $N$,
irrespective of $\alpha$.  Of course to calculate an isotherm, this
may also imply doubling the number of different $\rho$ at which
independent simulations must be performed.  Furthermore, as pointed
out by Stukan {\em et.\ al.\ }\cite{Stuk02} finite size effects become
more severe for larger polymer concentrations.  As an example, they
calculated the pressure of bond-fluctuation model polymers of length
$L=20$, using a box with a width $l_x=l_y=20$ and varying the distance
$l_z$ between the two hard walls used for thermodynamic
integration. At the rather high monomer packing fraction $\phi = N L/V
= 0.5$ they used, the influence of the two hard walls was noticeable
even up to distances of $l_z = 160$, where the RWTI method
overestimated the pressure by about $3\%$.  We performed comparable
simulations for $L=20$ SAW polymers at the same melt-like density.  In
a gravitational field we were able to reproduce the $l_z \rightarrow
\infty$ results of the RWTI method, but with the number of polymers
$N$ which the latter method would need for a size $l_z=20$ box only.  Thus the
less favourable scaling of our external field method with $N$ is
partially offset by less severe finite size effects, allowing
significantly fewer particles to be used.
\vspace*{-0.5cm}
\section{Conclusions}

We have shown that the hydrostatic equilibrium method, which allows
the e.o.s.\ of polymer solutions to be computed along an isotherm in a
single simulation, which determines the polymer concentration profiles
in a gravitational field, is both accurate and efficient.  It provides
the osmotic pressure as a function of concentration with a
computational effort which is a small fraction (typically under 5\%) of
that required using the RWTI method, since the latter requires a
series of MC simulations at different densities, and several
independent simulations at each density are needed for thermodynamic
integration.  Moreover the finite size problems which can affect
the RWTI method are insignificant in the present method, mainly
because the simulated system is ``open'' at high altitudes, {\it i.e.}\
essentially extends to infinity in the z-direction.  The hydrostatic
equilibrium method applies equally well to on and off-lattice models
of interacting polymers.  The only apparent limitation of the method
is that a large gravitational field, {\it i.e.}\ large values of
$\lambda_{cm}$ are required to achieve large densities.  For too large
$\lambda_{cm}$ the underlying LDA becomes inaccurate, and the inversion
procedure can lead to erroneous results.  Hence the method is expected
to be well adapted to dilute and semi-dilute solutions, while to reach
polymer densities typical of polymers melts, several runs with
increasing values of $\lambda_{cm}$ are needed to cover successive and
partially overlapping ranges of $\rho/\rho^*$.

We plan to extend the
present method to determine the osmotic e.o.s.\ of more complex
polymeric systems, including block copolymer solutions and melts.\\

CIA would like to thank the EPSRC for funding, AA is grateful for the support of the Royal Society.


\vspace*{-0.5cm}
\begin{thebibliography}{99}
\vspace*{-0.25cm}
\bibitem{Perc76} J.K. Percus, {\it J. Stat. Phys.} {\bf 1976}, {\it 15}, 423.
\bibitem{Bell73} A. Bellemans, E. Devos, {\it J. Polym. Sci. Symp.} {\bf 1973}, {\it 42}, 1195, R. Dickman and C.K. Hall, {\it J. Chem. Phys.} {\bf 1986}, {\it 85}, 3023.
\bibitem{Siep92} J.I.Siepmann, I.R. McDonald, D. Frenkel, {\it J. Phys.
Condens. Matter} {\bf 1992}, {\it 4}, 679.
\bibitem{Frenkelandsmit} D. Frenkel, B. Smit, ``Understanding Molecular
Simulation'', $2^{nd} Ed.$ Academic Press, New York, {\bf 1996}.
\bibitem{Dick87} R. Dickman, {\it J. Chem. Phys.} {\bf 1987}, {\it 87}, 2246.
\bibitem{Stuk02} M.R. Stukan, V.A. Ivanov, M. M\"uller, W. Paul, K. Binder,
{\it J. Chem. Phys.} {\bf 2002}, {\it 117}, 9934.
\bibitem{Bibe93} T. Biben, J.P. Hansen, J.L. Barrat, {\it J. Chem. Phys.}
{\bf 1993}, {\it 98}, 7330.
\bibitem{Dijk97} M. Dijkstra, J.P. Hansen, P.A. Madden, {\it Phys. Rev. E.}
{\bf 1997}, {\it 55} 3044.
\bibitem{Piaz93} R. Piazza, T. Bellini, V. Degiorgio, {\it Phys. Rev. Let.}
{\bf 1993}, {\it 71} 4267.
\bibitem{Gras95} P. Grassberger, R. Hegger, {\it J. Chem. Phys.} {\bf 1995}, {\it 102}, 6881.
\bibitem{Bolh01} P.G. Bolhuis, A.A. Louis, J.P. Hansen, E.J. Meijer, {\it J.
Chem. Phys.} {\bf 2001}, {\it 114}, 4296.
\bibitem{Addi04} C.I. Addison, A.A. Louis, J.P. Hansen, {\it J. Chem. Phys.}
{\bf 2004}, {\it 121}, 612.
\bibitem{Corey} R.H. Corey, M. Rubenstein, {\it Polymer Physics} (Oxford University Press, Oxford, {\bf 2003}

\end{thebibliography}
\end{document}